# All 2D Heterostructure Tunnel Field Effect Transistors: Impact of Band Alignment and Heterointerface Quality

*Keigo Nakamura[†], Naoka Nagamura[‡,§], Keiji Ueno[∥], Takashi Taniguchi[‡], Kenji Watanabe[‡], and Kosuke Nagashio[†*]*

[†]Department of Materials Engineering, The University of Tokyo, Tokyo 113-8656, Japan
[‡]National Institute for Materials Science, Ibaraki 305-0044, Japan,
[§]PRESTO, Japan Science and Technology Agency (JST), Saitama, 332-0012, Japan
[∥]Department of Chemistry, Saitama University, Saitama 338-8570, Japan
E-mail: nagashio@material.t.u-tokyo.ac.jp

**Abstract**
Van der Waals heterostructures are the ideal material platform for tunnel field effect transistors (TFETs) because a band-to-band tunneling (BTBT) dominant current is feasible at room temperature (RT) due to ideal, dangling bond free heterointerfaces. However, achieving subthreshold swing (SS) values lower than 60 mVdec$^{-1}$ of the Boltzmann limit is still challenging. In this work, we systematically studied the band alignment and heterointerface quality in $n$-MoS$_2$ channel heterostructure TFETs. By selecting a $p^+$-MoS$_2$ source with a sufficiently high doping level, stable gate modulation to a type III band alignment was achieved regardless of the number of MoS$_2$ channel layers. For the gate stack formation, it was found that the deposition of Al$_2$O$_3$ as the top gate introduces defect states for the generation current under reverse bias, while the integration of an $h$-BN top gate provides a defect-free, clean interface, resulting in the BTBT dominant current even at RT. All 2D heterostructure TFETs produced by combining the type III $n$-MoS$_2$/$p^+$-MoS$_2$ heterostructure with the $h$-BN top gate insulator resulted in low SS values at RT.

**INTRODUCTION**

The advanced metal-oxide-semiconductor field-effect transistor (MOSFET) technology now faces the dilemma for the reduction of power consumption because the subsequent reduction of power supply voltage leads to the increase in the leakage currents.[1] This fundamental limitation originates from the thermionic-based transport mechanism in MOSFETs, which prevents the subthreshold swing (SS), that is, steepness of the transfer characteristics in the subthreshold regime, from decreasing to less than 60 mVdec$^{-1}$ at room temperature (RT).[2] To overcome this limitation, two major concepts for steep-slope devices with SS values less than 60 mV/dec have been proposed: negative-capacitance (NC) FETs[3-5] and tunnel FETs (TFETs).[6-8] TFETs are more feasible because the device design rule based on the band-to-band tunneling (BTBT), that is, the high-energy tail of the Boltzmann distribution of carriers in the source is cut off by the band gap, is well established,[1,9,10] compared with the unclear physical picture in NC-FET.[11] In principle, however, TFETs suffer from a low drive current, which can be expected from the tunneling process. Recently, "bilayer" TFETs[12] have been studied as one of ideal device structures in some systems such as InGaAs/GaAsSb[13] and oxide/IV semiconductors,[14] where the effective tunneling area between the source and the channel is extended by the vertical stacking to increase the tunneling current. In exchange for the enhanced tunnel currents, the increased interface area of the bilayer structure increases the difficulty in controlling the interface quality, resulting in unsatisfactory SS values.

Two-dimensional (2D) materials are highly promising for use in bilayer TFETs with both high drive currents and low SS because the shorter tunneling distance and strong gate controllability can be expected from the van der Waals gap distances and the atomically sharp heterointerfaces that form independent of lattice matching and dangling bonds. Despite intensive research on many 2D TFET systems,[15-27] devices with SS values of sub-60 mVdec$^{-1}$ over several decades of drain current have rarely been demonstrated,[28] with the exception of devices utilizing ion gating with extremely high gate capacitance, however these types of devices are incapable of being integrated.[8,18,27] There are two main issues to overcome for 2D TFETs. One is the lack of the high doping sources and the other is the difficulty of controlling the interfacial properties.

There are few options for the high doping source materials required in the TFET device design since external doping techniques have not been developed.[29,30] The typical intrinsic high doping crystals like black phosphorus (BP) and SnSe$_2$ are not tolerant to oxidation and result in the formation of interlayer oxides.[24,31] Although local electrostatic doping in dual gate structures is often applied,[15,24,25,28] the structure of these types of devices becomes complicated and increases the parasitic capacitance unfavorably. Recently, by using a $p^+$-WSe$_2$ source that was doped by charge transfer from a WO$_x$ surface oxide layer, clear BTBT was successfully demonstrated in a stable $n$-MoS$_2$/$p^+$-WSe$_2$ TFET.[20] Although modulation from a type II (staggered gap) to a type III (broken gap) band alignment by only applying the gate bias was desired, an additional strong drain bias ($V_D$) was required to produce the type III band alignment suitable to the TFET operation. This unexpected result suggests that rigorous understanding of the band alignment of 2D/2D interfaces is required. According to the transmission probability calculated for carrier transport through the BTBT barrier,[1,9,10] both high-on and low-off currents are achievable by combining a source with a smaller energy gap ($E_G$) and a channel with a larger $E_G$. Although increasing $E_G$ values with decreasing the number of layers is a characteristic specific to 2D systems, how the band alignment changes as a function of the number of channel layers has not been systematically investigated despite its importance.

For the 2D/2D interfacial properties, the defect-free clean heterointerface[32] is critical for obtaining the BTBT dominant current under reverse bias at the diode. Although the BTBT current has been demonstrated at low temperatures, thermally activated behavior often appears at higher temperatures near RT.[15,19,20,26] That is, the generation current governs the total current, resulting in degradation of the SS at RT. This suggests that interface states exist even for 2D/2D interfaces. In general, high-$k$ top gate oxides have been used in most of 2D TFETs reported thus far to enhance the gate capacitance.[15,20,21,24] However, how the quality of the 2D/2D interface is affected by the deposition of high-$k$ oxides has not been revealed yet. Therefore, comparisons between high-$k$ and $h$-BN gate insulators should be carried out systematically in the same 2D TFET system,[33] because the use of $h$-BN in TFETs has been quite limited.

In this work, we systematically studied the band alignment in $n$-MoS$_2$ channel TFETs with two different sources, the charge transfer type $p^+$-WSe$_2$ and the substitutional type $p^+$-MoS$_2$. The doping level of $p^+$-MoS$_2$ was found to be higher than that of $p^+$-WSe$_2$, resulting in stable gate modulation to form the type III band alignment independent of the number of layers. Study of the gate stack revealed that the $h$-BN gate insulator retains the defect-free clean $p^+$-$n$ interface, unlike the deposition of high-$k$ oxide, resulting in the dominant BTBT current even at RT. All 2D heterostructure TFETs produced by combining the type III $n$-MoS$_2$/$p^+$-MoS$_2$ heterostructure with the $h$-BN top gate insulator achieved low SS values at RT.

## RESULTS

### Band alignment in $n$-MoS$_2$/$p^+$-WSe$_2$ heterostructures

**Figure 1** shows a schematic (a) and an optical micrograph (b) of the $n$-MoS$_2$/$p^+$-WSe$_2$ heterostructure on $h$-BN with an Al$_2$O$_3$ top gate insulator (30 nm) deposited by atomic layer deposition (ALD).[34,35] The $p^+$-WSe$_2$ FET was demonstrated by charge-transfer doping from the self-limiting WO$_x$ surface oxide layer, which was formed by ozone treatment.[36,37] This $p^+$-WSe$_2$ was further stabilized as the source crystal in TFETs by being transferred onto the $h$-BN substrate by using a polydimethylsiloxane (PDMS) sheet with an alignment system,[20] since the WO$_x$ layer was encapsulated by $h$-BN and WSe$_2$, as shown in the Supporting information of **Figure S1**. Typical thicknesses for the $h$-BN substrate and the $p^+$-WSe$_2$ source with the WO$_x$ layer were ~50 nm and ~40 nm, respectively. It should be noted that the thin $p^+$-WSe$_2$ (~15 nm) also showed similar results even though thick $p^+$-WSe$_2$ flakes were preferably used because of the easy transfer. For the $n$-MoS$_2$ channel, the number of layers was evaluated by measuring the optical contrast of $n$-MoS$_2$ on PDMS, as shown in the Supporting information of **Figure S2**. The $n$-MoS$_2$/$p^+$-WSe$_2$ heterostructure was prepared with the utmost care to the heterointerface quality by the dry PDMS transfer.[38,39] The formation of the clean heterointerface without any oxides is evident in the cross-sectional transmission electron microscopy (TEM) image of **Figure 1c**, which is one advantage of the stable WSe$_2$ crystal compared to other unstable, high doping crystals such as BP and SnSe$_2$.[24,31]

To determine the band alignment of the $n$-MoS$_2$/$p^+$-WSe$_2$ heterostructures with different numbers of $n$-MoS$_2$ channel layers, the diode properties were measured at 20 K at the top-gate voltage ($V_{TG}$) of 15 V, as shown in **Figure 1d**. The forward bias was placed on the positive side ($V_D$ is indeed negative), and vice versa for the reverse bias ($V_D$ is positive). It should be noted that the back-gate voltage ($V_{BG}$) was not applied in this study since the current was not improved by applying $V_{BG}$ and single gate structures are more feasible for real applications. Since the thermally excited carriers through the $p^+$-$n$ junction can be suppressed at low temperatures, the BTBT current from the valence band of $p^+$-WSe$_2$ to the conduction band of $n$-MoS$_2$ was clearly observed at the reverse bias side for all cases except for the 5 layer (L) $n$-MoS$_2$ channel. Moreover, we have already confirmed that the temperature dependence of the current expected at the source/drain contacts is negligible due to ohmic contact,[20] indicating that the voltage is applied predominantly to the $p^+$-$n$ junction. The diode properties firmly reflect the band alignment of the $n$-$MoS_2$/$p^+$-$WSe_2$ heterostructures under sufficiently large $V_{TG}$. Interestingly, the layer number dependence is evident, where the band alignment under zero $V_D$ changes from type II at 5L to type III at 3L and again to type II at 1L, as schematically illustrated at the bottom of **Figure 1d**.

Before considering how the layer number dependence affects the band alignment, let us focus on the 3L $n$-$MoS_2$ channel with the type III band alignment, which is suitable for the TFET operation. **Figure 2a** shows the full set of diode properties at different $V_{TG}$ at 20 K. For $V_{TG}$ = -12 V, the $MoS_2$ channel is slightly $n$-type. The diffusion current for the forward bias and the noise level of the saturation current for the reverse bias are present, which clearly shows the rectified behavior. Hence, the band alignment is type II with a large band overlap, which is schematically the same as "5L" shown in **Figure 1d**. For -12 V < $V_{TG}$ < -9 V, the $MoS_2$ channel is $n$-type. The reverse BTBT currents are enhanced with increasing $V_{TG}$. The band alignment under zero $V_D$ is still type II but the band overlap becomes smaller, which is schematically the same as "4L" shown in **Figure 1d**. Finally, for $V_{TG}$ > -8 V, the $MoS_2$ channel is highly $n$-type. A negative differential resistance (NDR) trend formed by the transition from the BTBT current to the diffusion current is evident for the forward bias, which indicates that the band alignment is type III. These results clearly indicate that the band alignment of the $n$-$MoS_2$/$p^+$-$WSe_2$ heterostructure under zero $V_D$ can be readily changed from type II to type III by only using the $V_{TG}$. When the temperature is increased with the constant $V_{TG}$ = 15 V, the NDR trend gradually disappears at ~160 K since thermally excited carriers are generated at the $p^+$-$n$ junction, as shown in **Figure 2b**. The Arrhenius plot of the current at the reverse bias of -2 V in **Figure 6e** suggests thermally activated behavior at high temperatures and temperature-independent behavior at low temperatures, which also supports the presence of the BTBT current at low temperatures.

**Figure 2c** shows the transfer characteristics of another TFET device at the reverse bias of -2 V at different temperatures. The SS values were estimated as a function of the drain current ($I_D$) in **Figure 2d**. The SS values increase with increasing $I_D$, which is consistent with the typical SS behavior of TFETs. As can be expected from the thermally activated behavior of the current, the SS values are constant in the low temperature range and start to increase with increasing temperature. Although the clean 2D/2D interface was achieved, the lowest SS values obtained at RT and 20 K are 358 mVdec$^{-1}$ and 106 mVdec$^{-1}$, respectively.

It is necessary to discuss how the dependence of the band alignment on the number of $MoS_2$ layers, which is schematically illustrated at the bottom of **Figure 1d**. The BTBT onset voltage ($V_{BTBT}$) is defined as the voltage at 10$^{-11}$ A in the diode property and is scaled to the band offset between the conduction band minimum (CBM) of $MoS_2$ and the valance band maximum (VBM) of $WSe_2$, as shown in the Supporting information of **Figure S3**. **Figure 3a** shows the $V_{BTBT}$ as a function of $V_{TG}$ for different numbers of $MoS_2$ layers in the $n$-$MoS_2$/$p^+$-$WSe_2$ heterostructure. The 5L $MoS_2$ channel is excluded from this figure since BTBT was not observed, as shown in **Figure 1d**. When the $V_{BTBT}$ reaches 0 V by applying the $V_{TG}$, the band alignment changes from type III to type II. Interestingly, the $V_{BTBT}$ for the 3L $MoS_2$ channel can only reach 0 V (type III), while the $V_{BTBT}$ for different numbers of $MoS_2$ layers saturates before reaching 0 V even under sufficiently large $V_{TG}$ (type II). Here, the dependence of $E_G$ on the number of $MoS_2$ layers may be expected as one of origins for this layer number dependence on the band alignment. According to first-principles calculations,[40] the VBM increases rapidly as the number of layers increases with respect to the vacuum level, while the CBM remains almost constant. Since the band alignment between the CBM for $n$-$MoS_2$ and the VBM for $p^+$-$WSe_2$ is being considered, the dependence of $E_G$ on the number of $MoS_2$ layers does not explain the observed band alignment.

Looking back at the saturation behavior of $V_{BTBT}$ that suggests that Fermi level ($E_F$) in $MoS_2$ is sufficiently modulated into the conduction band, it can be considered that the restriction to the type II band alignment is due to $WSe_2$, not $MoS_2$. That is, the $E_F$ of $WSe_2$ is moved into the $E_G$ from inside of the valence band due to a reduction in $p$-doping, as schematically shown in **Figure 3b**. In this case, the type III band alignment will never be realized. The reduction of $p$-doping in $WSe_2$ can originate from two different sources. One is electron transfer from $MoS_2$ to $WSe_2$ and the other is that the top gate modulates $WSe_2$ as well as $MoS_2$. The $V_{BTBT}$ values at $V_{TG}$ = 15 V from **Figure 3a** are plotted as a function of the number of $MoS_2$ layers in **Figure 3c**, which suggests that the band alignment may be controlled by two different physical origins since the upward convex behavior is not readily explained by a single physical property. When $p^+$-$WSe_2$ is in contact with $MoS_2$, electrons are transferred from $MoS_2$ to $WSe_2$, as illustrated in **Figure 3d**, and the $E_F$ of $WSe_2$ will increase uniformly because $p^+$-$WSe_2$ is formed by the transfer of electrons to $WO_x$. Because the amount of transferred electrons increases with the number of MoS$_2$ layers, the $E_F$ of WSe$_2$ also changes with the number of MoS$_2$ layers and the band alignment becomes type II for the 4L and 5L MoS$_2$ channels, which is indicated by the blue line in **Figure 3c**. On the other hand, when the number of the MoS$_2$ layers is reduced from 5L to 1L, WSe$_2$ is also expected to be modulated by the top gate. Therefore, by applying $V_{TG}$, the $E_F$ of WSe$_2$ and MoS$_2$ are both modulated at the same time, which reduces the rate of change of the $V_{BTBT}$ for the 1L and 2L MoS$_2$ channels in **Figure 3a**. Therefore, the band alignment changes from type III to type II as the number of MoS$_2$ layers decreases, as shown by the orange line in **Figure 3c**. Because of this, only the 3L MoS$_2$ channel appears as type III due to the combination of the two different sources of $p$-doping reduction for the WSe$_2$.

Although experimentally confirming modulation of the $E_F$ of $p^+$-WSe$_2$ by the top gate is difficult, it is possible to prove there is electron transfer from MoS$_2$ to $p^+$-WSe$_2$. Here, a core-level photoelectron spectromicroscopic apparatus installed at the synchrotron radiation facility of SPring-8, called "3D nano-ESCA,"[41,42] is utilized, with which we can scan a sample with a high lateral spatial resolution of ~70 nm to record photoelectron spectra for quantitative analysis of the chemical states. **Figure 4a** shows an intensity mapping image for the Mo 3d$_{5/2}$ peak with a spatial resolution of 200 nm for the 3L-$n$-MoS$_2$/$p^+$-WSe$_2$ heterostructure on $h$-BN specifically prepared for 3D nano-ESCA, as shown in the Supporting information of **Figure S4**. The position of the heterostructure is clearly identified. **Figure 4b** shows the pinpoint core-level spectra for the W 4f$_{5/2}$ and 4f$_{7/2}$ peaks recorded at points (i) $p^+$-WSe$_2$/$h$-BN and (ii) 3L-$n$-MoS$_2$/$p^+$-WSe$_2$/$h$-BN. The W 4f peaks for (ii) where $p^+$-WSe$_2$ is in contact with 3L-$n$-MoS$_2$ shifts to higher binding energy than that for point (i) where $p^+$-WSe$_2$ is directly on $h$-BN. When $p^+$-WSe$_2$ is doped with electrons, the $E_F$ of $p^+$-WSe$_2$ increases, resulting in the W 4f peaks having higher binding energy. These results support that $p$-doping reduction is occurring in WSe$_2$.

**Realization of stable type III band alignment using a $p^+$-MoS$_2$ source**
In the TFET transfer characteristics, the on-current is one of important figures of merit. However, the on-current in the type II band alignment is much lower than that for the type III alignment, which can be realized at the current level at the reverse bias of -2 V in **Figure 1d** (also in the Supporting information of **Figure S5**). As discussed in the previous section, the $E_F$ of the WSe$_2$ source was modulated by the surrounding, which prevented from controlling the band alignment from type II to type III. The problem for this is that the doping level of $p^+$-WSe$_2$ is not sufficiently high. To mitigate this shortcoming, a niobium (Nb)-doped $p^+$-MoS$_2$ crystal[43,44] was used as an alternative source. Compared to the charge transfer type $p^+$-WSe$_2$, the Nb-doped $p^+$-MoS$_2$ is a thermodynamically stable substitutional type source with a degenerate hole concentration of ~3×10$^{19}$ cm$^{-3}$, which was confirmed by Hall measurement.[43] **Figure 5** compares the transfer characteristics of (a) $p^+$-WSe$_2$ and (b) $p^+$-MoS$_2$ FETs at different temperatures. The linear scale is also shown in the Supporting information of **Figure S6**. No gate dependence was observed for the transfer characteristics of the $p^+$-MoS$_2$ FET for all the measured temperatures compared with the slight change for the $p^+$-WSe$_2$ FET. This indicates that the doping level of $p^+$-MoS$_2$ is higher than $p^+$-WSe$_2$. The channel thickness dependence of the transfer characteristics of $p^+$-MoS$_2$ is also available in our previous research,[33] where it was revealed that the narrow maximum depletion width of ~7 nm is consistent with the degenerate hole concentration. Indeed, when $p^+$-MoS$_2$ was used as the source in the heterostructure TFET with the MoS$_2$ channel, the diode properties in **Figure 5c** indicate that the type III band alignment was achieved even for the 1L $n$-MoS$_2$ channel as well as the 3L $n$-MoS$_2$ channel at $V_{TG}$ = 0 V. The successful modulation of the band alignment to type III by only gate modulation supports the above conclusion that the reduction in the $p$-doping of WSe$_2$ is indeed a limiting factor that restricts the band alignment to type II.

**Demonstration of low SS values for All 2D heterostructure TFETs**
We demonstrated that type III band alignment can be obtained for MoS$_2$ channels with any number of layers using the $p^+$-MoS$_2$ source. However, the lowest SS values obtained for both the $p^+$-MoS$_2$ source and the $p^+$-WSe$_2$ source were restricted to 137 mVdec$^{-1}$, as shown in the Supporting information of **Figure S5**. There are three strategies to further reduce the SS values: (i) Recently, we have discovered that the deposition of Al$_2$O$_3$ top gate oxide on the monolayer MoS$_2$ channel on $h$-BN substrate increases the interface states density at the "bottom" MoS$_2$/$h$-BN interface as well as the top Al$_2$O$_3$/MoS$_2$ interface due to the introduction of strain in the MoS$_2$ channel.[33] This suggests that the 2D/2D interface in TFET is also degraded by the high-$k$ deposition. Therefore, an $h$-BN top gate insulator was adopted to benefit from the electrically inert interface in 2D heterostructure TFETs.[32] (ii) The $p^+$-MoS$_2$ source was used instead of the $p^+$-WSe$_2$ source because the $E_F$ of $p^+$-MoS$_2$ cannot be modulated due to the degenerately high doping of the $p^+$-MoS$_2$. (iii) According to the transmission probability calculated for carrier transport through the BTBT barrier,[1,9,10] the $E_G$ for the channel should be larger than that for the source to keep the off current low but $E_G$ also should be as small as possible to increase the transmission probability. Therefore, the 1L and 3L MoS$_2$ channels were compared. Based on these three considerations, all 2D heterostructure TFETs were fabricated to achieve SS values lower than 60 mVdec$^{-1}$.

**Figure 6** shows a schematic (a) and an optical micrograph (b) of a typical $h$-BN/$n$-MoS$_2$/$p^+$-MoS$_2$/$h$-BN all 2D heterostructure TFET. The typical thickness for the top gate $h$-BN insulator and the $p^+$-MoS$_2$ source are ~15 nm and ~30 nm, respectively. The atomically sharp gate stack interfaces are clearly seen in the cross-sectional TEM image of **Figure 6c** since all of the 2D materials are stable in air. As was expected, the diode properties of the all 2D heterostructure TFET with the 3L-$n$-MoS$_2$ channel in **Figure 6d** shows the type III band alignment at $V_{TG}$ = 6 V. The NDR trend at the forward side is not visible, unlike the case in **Figure 2b**. This may be explained by the two possibilities. One is large energy gap between VBM for $p^+$-MoS$_2$ and CBM for $n$-MoS$_2$ in the type III band alignment because $E_F$ in $p^+$-MoS$_2$ is located deeply in the valence band due to higher doping concentration. The other is the suppression of the diffusion current due to the larger barrier between the CBM for $p^+$-MoS$_2$ and the VBM for the $n$-MoS$_2$ channel because the $E_G$ of bulk MoS$_2$ (~1.4 eV) is larger than the $E_G$ of bulk WSe$_2$ (~1.2 eV). An Arrhenius plot of the current at the reverse bias of -2 V is compared with other heterostructures in **Figure 6e**. It should be noted that all four heterostructure TFETs exhibit type III band alignment. For the $h$-BN top gate heterostructure devices with the 1L and 3L MoS$_2$ channels, temperature-independent behavior is evident over the entire temperature range, indicating that BTBT is dominant even at RT and that the source/drain contacts are Ohmic nature. This is quite promising for TFET operation with low SS values at RT. On the other hand, when Al$_2$O$_3$ was used as the top gate insulator, thermally activated behavior at high temperatures was clearly observed regardless of the source crystal. These comparisons indicate that the trap-related generation-recombination current[45] and/or the trap-assisted tunneling current[46,47] under reverse bias are drastically suppressed by the successful integration of the electrically inert interface in the 2D heterostructure TFET.

Finally, the transfer characteristics of the 2D heterostructure TFETs at the reverse bias of -2 V at RT are shown in **Figure 6f**. According to our previous studies on the dielectric breakdown of $h$-BN,[48,49] the vertical dielectric breakdown field is ~1.2 V/nm and the leakage occurs at ~0.5 V/nm. Therefore, in this study, the $V_{TG}$ sweep range for the $h$-BN top gate is determined based on the critical electrical field of ~0.4 V/nm. Compared with the exact control of Al$_2$O$_3$ thickness, the thickness of $h$-BN cannot be well controlled. Therefore, the $V_{TG}$ sweep range differs device to device. The estimated SS values are shown as a function of $I_D$ in **Figure 6g**. SS values for all 2D heterostructure TFETs are much lower than that for 2D heterostructure TFET with Al$_2$O$_3$ top gate. Moreover, in all 2D heterostructure TFETs, the smaller $E_G$ of the 3L-$n$-MoS$_2$ channel was preferable, compared with the 1L-$n$-MoS$_2$ channel. Here, let us analyze the best device, 3L-$n$-MoS$_2$/$p^+$-MoS$_2$ heterostructure, more in detail. The leakage current contributions should be considered carefully since artificially low SS values are often reported, as shown in the Supporting information of **Figure S6**. It is evident that $I_D$ overlaps with $I_S$ for the 3L-$n$-MoS$_2$ channel because there is no gate leakage (**Figure S6c**), which supports that the SS value is not artificial. However, the hysteresis was detected in the transfer characteristics and SS even for all 2D heterostructure TFET, as shown in **Figure 6f and 6g**. In general, the existence of hysteresis results in the degraded SS value and never provides the "apparently" better SS value. However, for the actual device applications, SS values in both sweep directions should be low enough for 4-5 decade of the current. By improving the cleanness of the 2D heterointerface, for example, removing the bubbles, further reduction of the hysteresis and the SS value will be possible.

Recently, quite low average SS values over 4 dec of current (SS = ~22.9 mVdec$^{-1}$) have been reported for "in-plane type" BP/BP heterostructure TFETs,[28] which are different from the "bilayer structure" (out-of-plane) TFETs reported in this study. Density function theory (DFT) calculations[50] suggest that out-of-plane 2D/2D heterostructures are more preferable than in-plane 2D/2D heterostructures in terms of gate controllability because the dangling bond states remain even after the defect free heterointerface is formed for the calculation, as schematically shown in the Supporting information of **Figure S7**. The fact that such a low SS value was obtained for the in-plane TFETs suggests the lower SS values are possible for out-of-plane TFETs with higher on-currents when the whole heterointerface is more rigorously controlled by ensuring the interface is clean and the lattices match.

**CONCLUSIONS**
Band alignment and interfacial quality were critical for achieving SS values lower than 60 mVdec$^{-1}$ of the Boltzmann limit. Insufficient doping levels in the source crystal restricted the band alignment to type II, even under sufficient gate bias. This strongly suggests the importance of establishing external doping techniques for the success of future studies. The key finding regarding the quality of the heterointerface is that producing the defect-free clean heterointerface via integration of the *h*-BN top gate provides the BTBT dominant current even at RT. All 2D heterostructure TFETs produced by combining the type III *n*-MoS$_2$/*p*$^+$-MoS$_2$ heterostructure with the *h*-BN top gate insulator resulted in low SS values at RT. Since it has been suggested that out-of-plane 2D/2D heterostructures are more preferable than in-plane 2D/2D heterostructures in terms of gate controllability, further reductions in SS values and higher on-currents are possible when the entire heterointerface is more rigorously controlled.

**EXPERIMENTAL METHODS**
**Device fabrication.** Natural *n*-MoS$_2$ and Nb-doped *p*$^+$-MoS$_2$ bulk crystals were purchased from SPI Supplies and HQ graphene, respectively, whereas WSe$_2$ and *h*-BN bulk crystals were grown using a physical vapor transport technique without an I$_2$ transport agent[51] and a temperature-gradient method under a high-pressure and high-temperature atmosphere.[52] The thin 2D layers were prepared from the mechanical exfoliation of the bulk crystals. The *n*-MoS$_2$/*p*$^+$-WSe$_2$ heterostructure on the *h*-BN substrate was fabricated using a dry transfer method with PDMS and an alignment system.[38,39] Ni/Au was deposited as the source/drain electrodes after the electrode pattern was formed using electron beam lithography. For the high-*k* top-gate formation, 1-nm-thick Y$_2$O$_3$ was deposited via thermal evaporation of Y metal in a PBN crucible in an Ar atmosphere with a partial pressure of $10^{-1}$ Pa to form a buffer layer.[34] Al$_2$O$_3$ oxide layers with thicknesses of 30 nm were deposited via ALD, followed by formation of the Al top-gate electrode.[35] Alternatively, *h*-BN top gate TFETs were also fabricated by the dry transfer method. No additional annealing was performed after fabrication of the heterostructures.

**Measurements.** Raman spectroscopy and AFM were employed to determine the crystal quality and thickness of the flakes. TEM images were taken at an acceleration voltage of 200 kV using a JEM-ARM200F to confirm the quality of the 2D heterostructure interface. For core-level photoelectron spectromicroscopy measurements, a 3D nano-ESCA installed at the soft X-ray beamline BL07LSU in the synchrotron radiation facility of SPring-8 was used. The photon energy of the incident beam was 1000 eV. All electrical measurements were performed using a Keysight B1500 in a vacuum prober with a cryogenic system.

**SUPPORTING INFORMATION**
The up-side-down transfer, optical contrast of MoS$_2$, diode property of type II band alignment, 3D nano-ESCA, Transfer characteristics of various heterostructure TFETs, analysis on the leakage current, band alignment of out-of-plane and in-plane heterojunction. The Supporting Information is available free of charge via the Internet at http://pubs.acs.org.

**AUTHOR INFORMATION**
**Corresponding Author**
*Email: nagashio@material.t.u-tokyo.ac.jp

**Notes**
The authors declare no competing financial interests.

**ACKNOWLEDGEMENTS**
This research was supported by Samco Science and Technology Foundation, The Canon Foundation, the Elemental Strategy Initiative conducted by the MEXT, Grant Number JPMXP0112101001, the JSPS Core-to-Core Program, A. Advanced Research Networks, the JSPS A3 Foresight Program, JSPS KAKENHI Grant Numbers JP20H00354, JP19H00755, 19K21956, 18H03864, and 19H02561, and CREST(Grant number: JPMJCR15F3) and PRESTO (Grant number: JPMJPR17NB) commissioned by the Japan Science and Technology Agency (JST), Japan. The spectral datasets were obtained with the support of the University of Tokyo outstation beamline at SPring-8 (Proposal Numbers: 2018B7580 and 2019A7451).

**FIGURES**

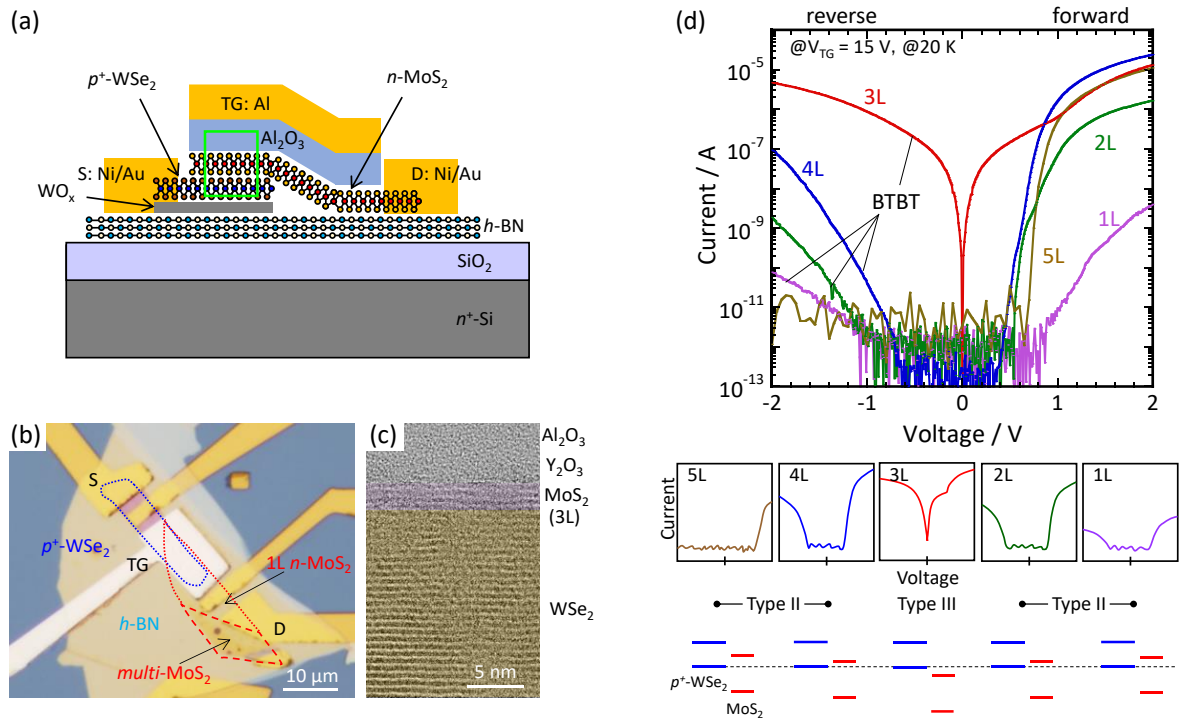

**Figure 1** a) Schematic illustration and b) optical micrograph of the $n$-MoS$_2$/$p^+$-WSe$_2$ heterostructure on $h$-BN with an Al$_2$O$_3$ top gate insulator. c) Cross-sectional TEM image of the Al$_2$O$_3$/3L-$n$-MoS$_2$/$p^+$-WSe$_2$ heterostructure from the solid rectangular in a). The number of MoS$_2$ layers is 3. d) Diode properties for the $n$-MoS$_2$/$p^+$-WSe$_2$ heterostructure with different numbers of MoS$_2$ layers at $V_{TG}$ = 15 V and 20 K. The layer number dependent characteristics are schematically illustrated at the bottom.

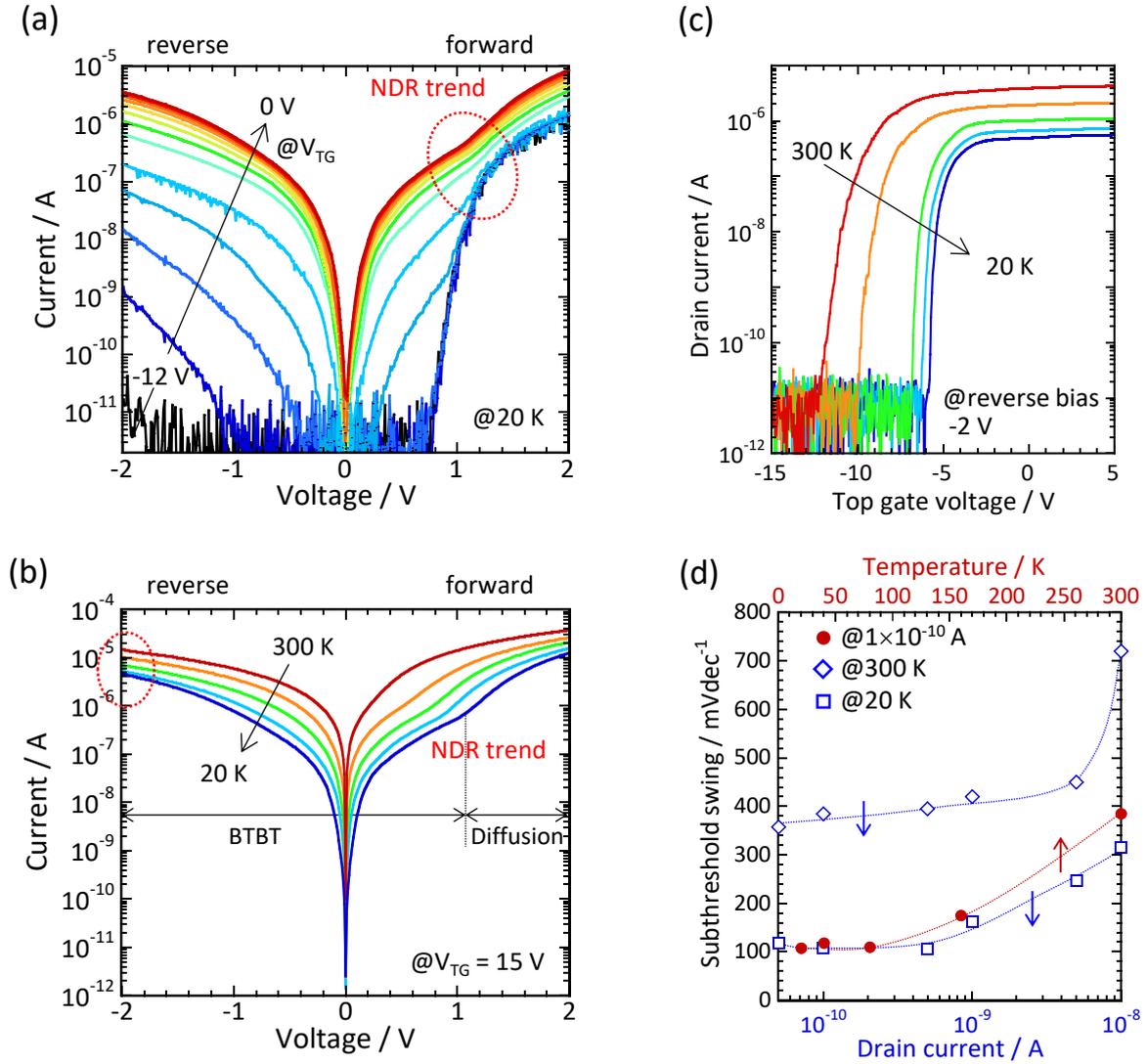

**Figure 2** a) Diode properties for the 3L-*n*-MoS$_2$/$p^+$-WSe$_2$ heterostructure at different $V_{TG}$ and 20 K. The voltage step in $V_{TG}$ is 1 V. b) Temperature dependence of the diode properties for the 3L-*n*-MoS$_2$/$p^+$-WSe$_2$ heterostructure at a constant $V_{TG}$ = 15 V (20, 40, 80, 160, and 300 K). c) Transfer characteristics of the TFET at different temperatures under a reverse bias of -2 V. d) SS as a function of $I_D$ and the temperature.

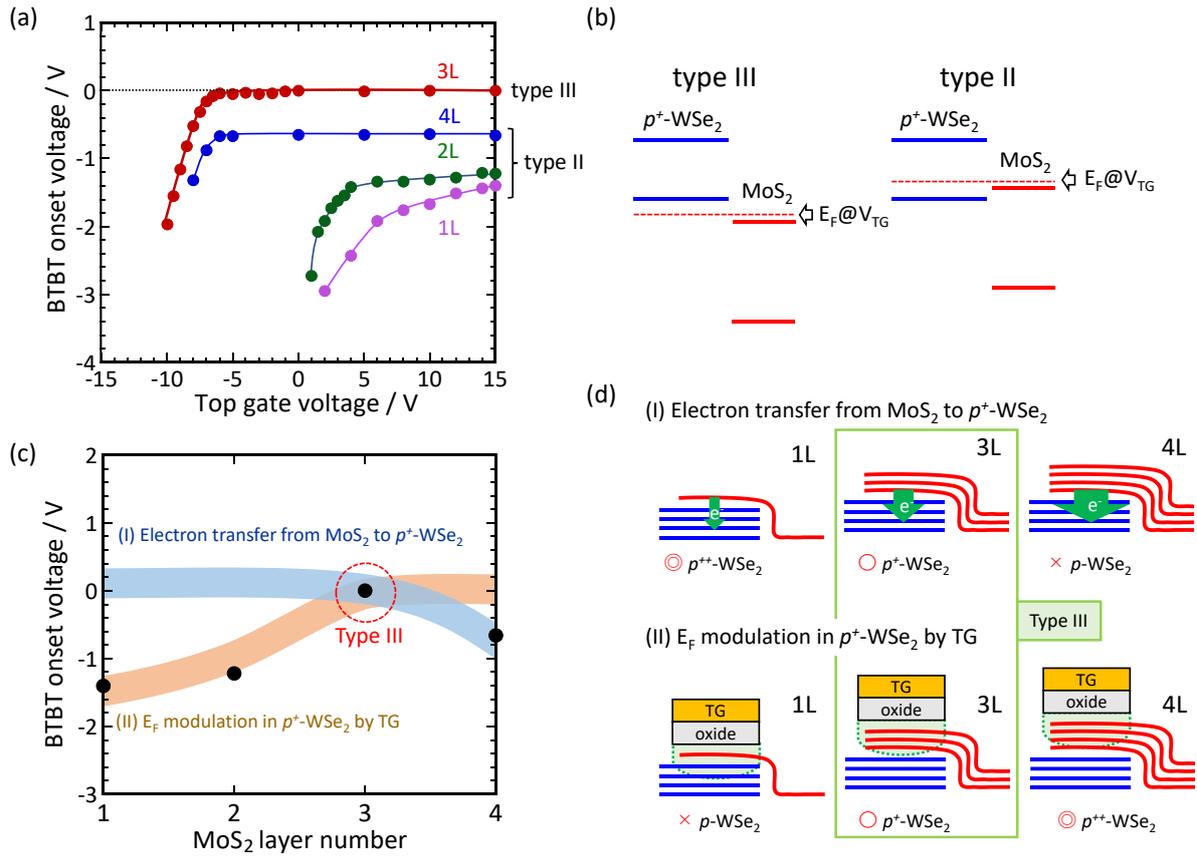

**Figure 3** a) $V_{BTBT}$ as a function of $V_{TG}$ for MoS$_2$ with different numbers of layers the 3L-$n$-MoS$_2$/$p^+$-WSe$_2$ heterostructure. b) Schematic illustration of the band alignment for type II and type III. $E_F$ for MoS$_2$ is fixed in the conduction band. When the $E_F$ of WSe$_2$ is in the valence band, the band alignment can be type III. On the other hand, when the $E_F$ of WSe$_2$ is in the band gap, it is restricted to type II. c) $V_{BTBT}$ as a function of the number of MoS$_2$ layers. d) Schematic illustration of the two different physical origins for the $p$-doping reduction of WSe$_2$.

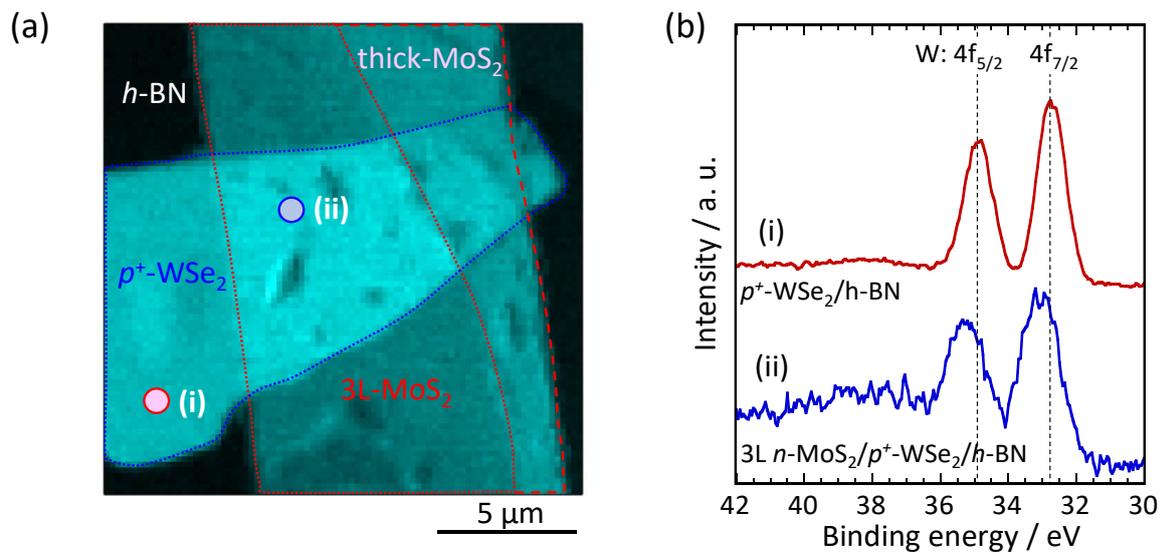

**Figure 4** a) Photoelectron intensity mapping image for the Mo $3d_{5/2}$ peak with a spatial resolution of 200 nm for the 3L-$n$-MoS$_2$/$p^+$-WSe$_2$ heterostructure on $h$-BN. b) Pinpoint core-level spectra for W $4f_{5/2}$ and $4f_{7/2}$ peaks recorded at points (i) and (ii).

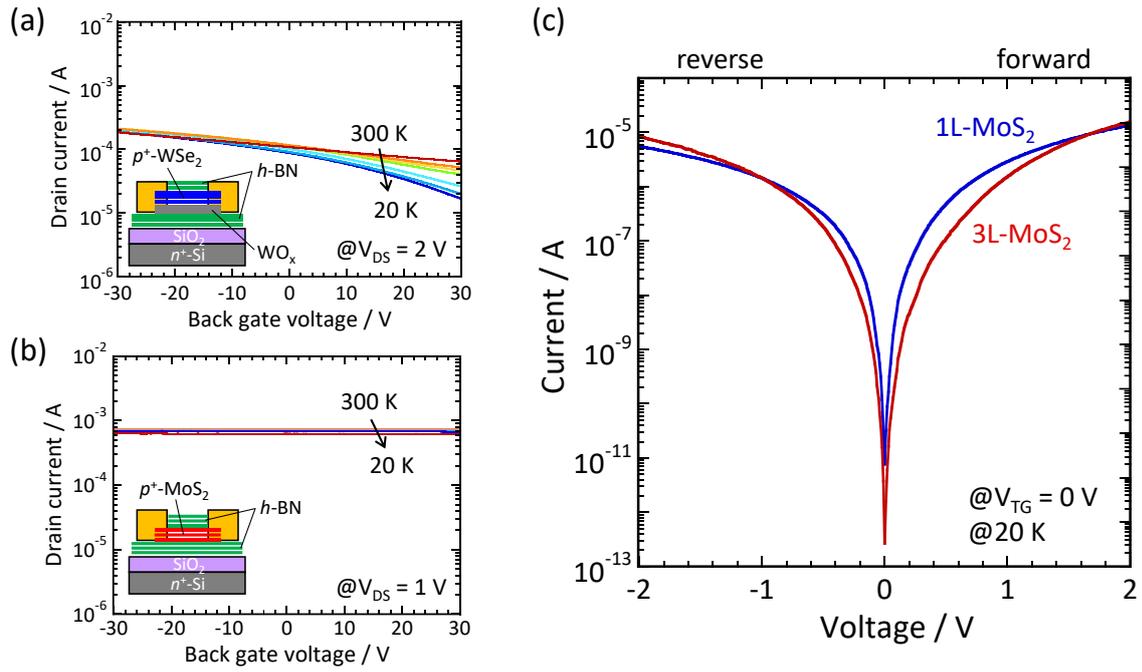

**Figure 5** Transfer characteristics for a) $p^+$-WSe$_2$ FET and b) $p^+$-MoS$_2$ FET at different temperatures (20, 40, 80, 160, 200, 250, and 300 K). c) Diode properties for 3L-$n$-MoS$_2$/$p^+$-MoS$_2$ and 1L-$n$-MoS$_2$/$p^+$-MoS$_2$ heterostructures at $V_{TG}$ = 0 V at 20 K. Both structures are the same as that in **Figure 1a**, except for $p^+$-MoS$_2$.

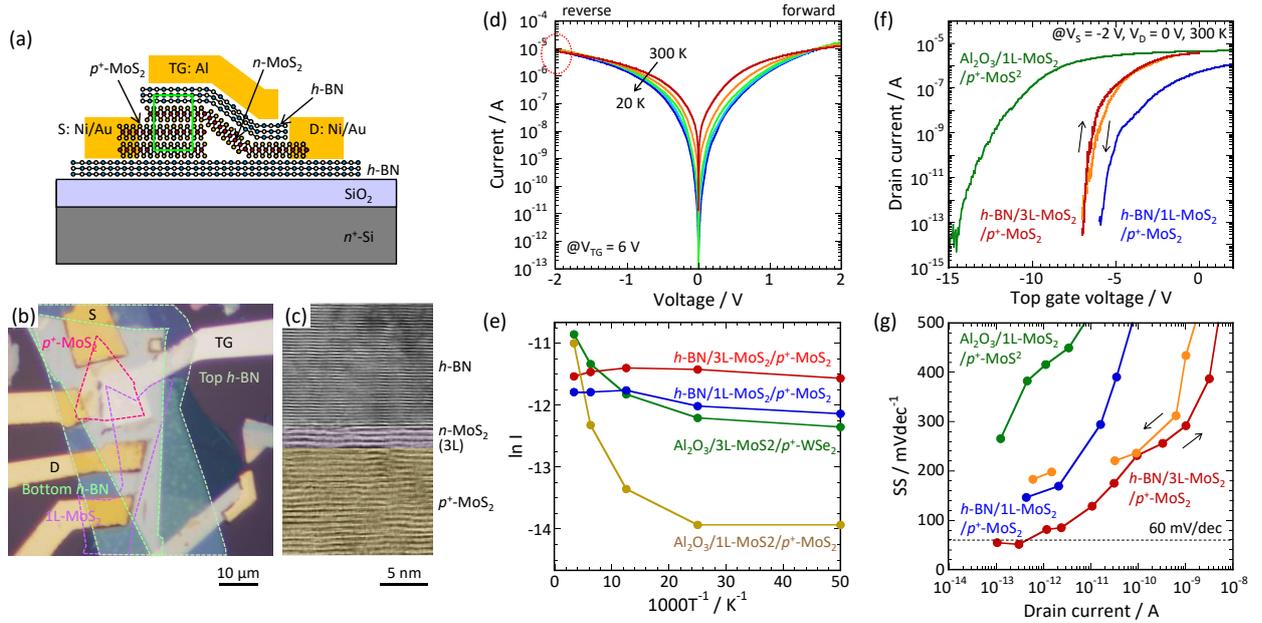

**Figure 6** a) Schematic illustration and b) optical micrograph of all 2D heterostructure TFET. c) Cross sectional TEM image of all 2D heterostructure at the solid rectangular in a). The number of MoS$_2$ layers is 3. d) Diode properties in the 3L-$n$-MoS$_2$/$p^+$-MoS$_2$ heterostructure at $V_{TG}$ = 6 V and different temperatures (20, 40, 80, 160, and 300 K). e) Arrhenius plot of the current at the reverse bias of -2 V for different heterostructures. f) Transfer characteristics for the three different heterostructure TFETs. Round sweep behavior is shown only for 3L-$n$-MoS$_2$/$p^+$-MoS$_2$ heterostructure (red & orange). g) SS as a function of $I_D$ for the three different heterostructure TFETs. Round sweep behavior is shown only for 3L-$n$-MoS$_2$/$p^+$-MoS$_2$ heterostructure (red & orange).

**TOC**

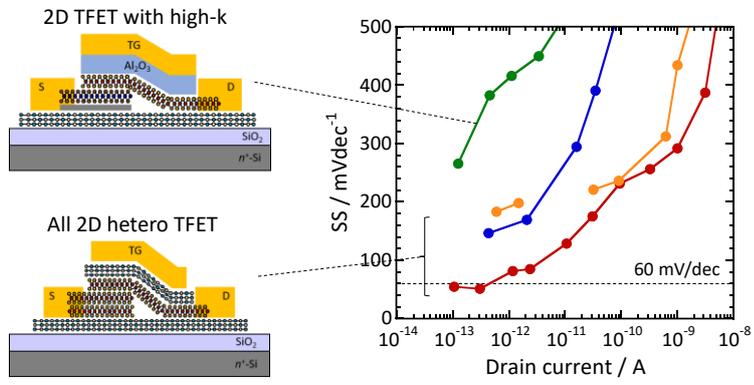

**Supporting information**

# All 2D heterostructure Tunnel Field Effect Transistors: Impact of band alignment and heterointerface quality

*Keigo Nakamura[†], Naoka Nagamura[‡,§], Keiji Ueno[∥], Takashi Taniguchi[‡], Kenji Watanabe[‡], and Kosuke Nagashio[†*]*

[†]Department of Materials Engineering, The University of Tokyo, Tokyo 113-8656, Japan

[‡]National Institute for Materials Science, Ibaraki 305-0044, Japan,

[§]PRESTO, Japan Science and Technology Agency (JST), Saitama, 332-0012, Japan

[∥]Department of Chemistry, Saitama University, Saitama 338-8570, Japan

**E-mail:** nagashio@material.t.u-tokyo.ac.jp

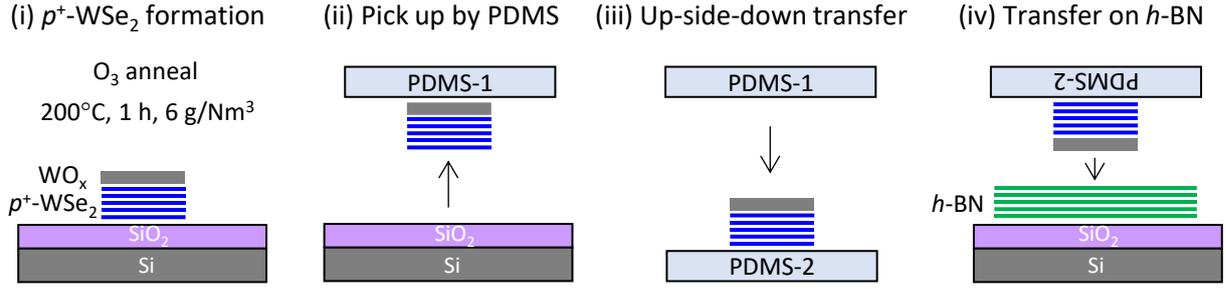

**Figure S1** Schematic drawing for the procedures of (i)~(iv) to stabilize $p^+$-WSe$_2$.

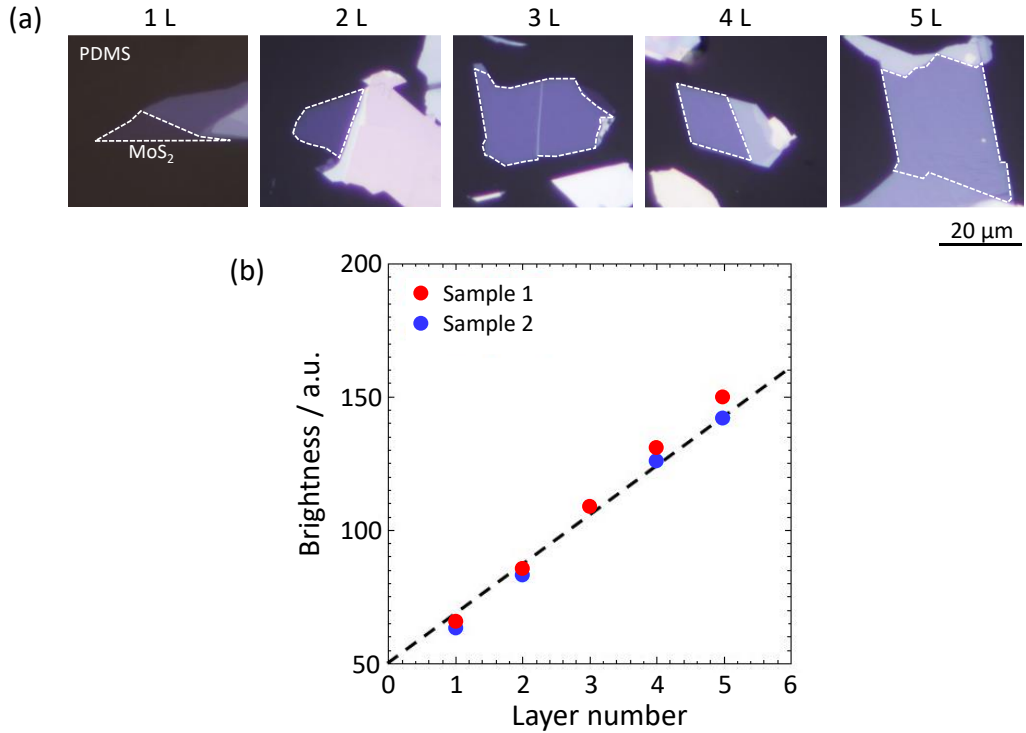

**Figure S2** (a) Optical images of MoS$_2$ from monolayer to 5 layers on PDMS.   (b) Brightness of MoS$_2$ on PDMS as a function of layer number. For the brightness calibration, the brightness of PDMS was initially adjusted to be 50 in 256 by changing the optical light intensity. Then, the brightness of MoS$_2$ on PDMS was measured. The MoS$_2$ layer number was confirmed by Raman and AFM measurements. These brightness values are plotted for two different samples. Moreover, in case of $h$-BN on PDMS, the brightness for 27-nm $h$-BN is found to be 85. The thickness of $h$-BN is roughly identified before the transfer and measured by AFM after the heterostructure formation.

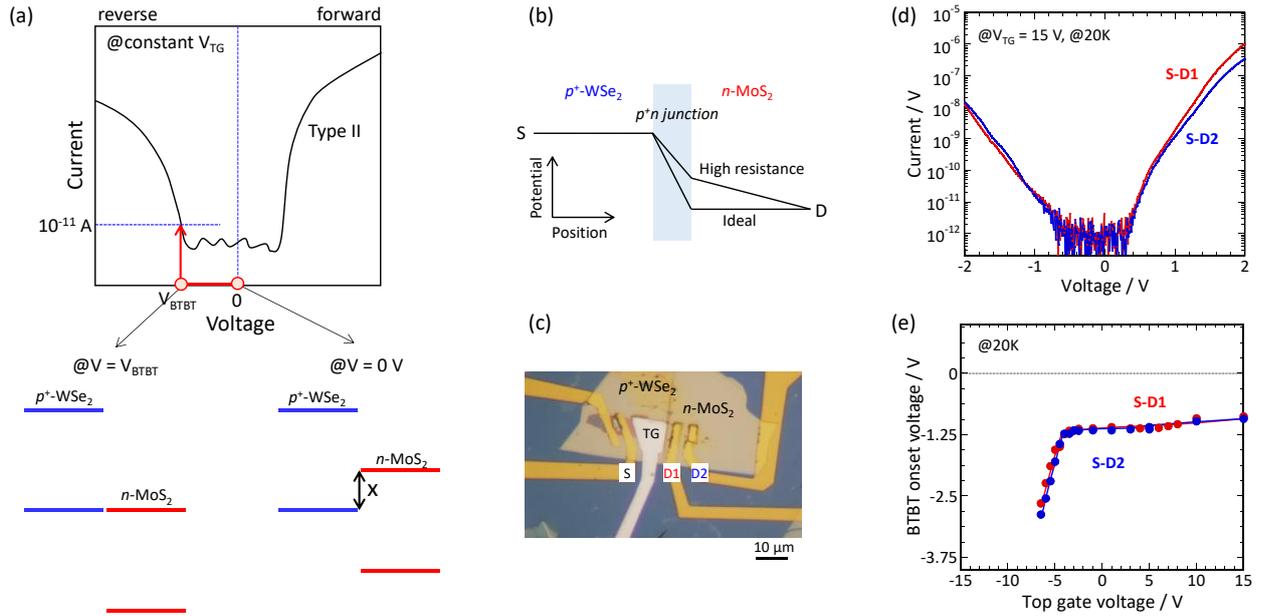

**Figure S3** Schematic illustration of the diode properties of the type II band alignment at the constant $V_{TG}$ is shown in (a). $V_{BTBT}$ is defined as the voltage at $10^{-11}$ A. The band alignment @$V$ = 0 V is type II with band overlap. When $V_{BTBT}$ is applied, the band alignment becomes type III in which the energy level of the top of the valence band of $p^+$-WSe$_2$ is consistent with that of the conduction band minimum for $n$-MoS$_2$. Therefore, $V_{BTBT}$ is scaled to the band offset of "x". However, in order that $V_{BTBT}$ implies the band offset, all the voltage drop must be consumed only at the $p^+$-$n$ junction, not at the MoS$_2$ channel, as shown in (b). Therefore, the $p^+$-WSe$_2$/4L-$n$-MoS$_2$ heterostructure with multi terminals were additionally fabricated in (c). Diode properties were measured using different terminals (S-D1 & S-D2) at $V_{TG}$ = 15 V and 20 K as shown in (d). $V_{BTBT}$ is also plotted as a function of $V_{TG}$ in (e). It is evident that there is no difference in $V_{BTBT}$ between S-D1 and S-D2, supporting that there is no voltage drop in the access region of the $n$-MoS$_2$ channel. Since the voltage drop at the contact has been neglected due to the Ohmic contact, it can be said that $V_{BTBT}$ implies there is band offset.

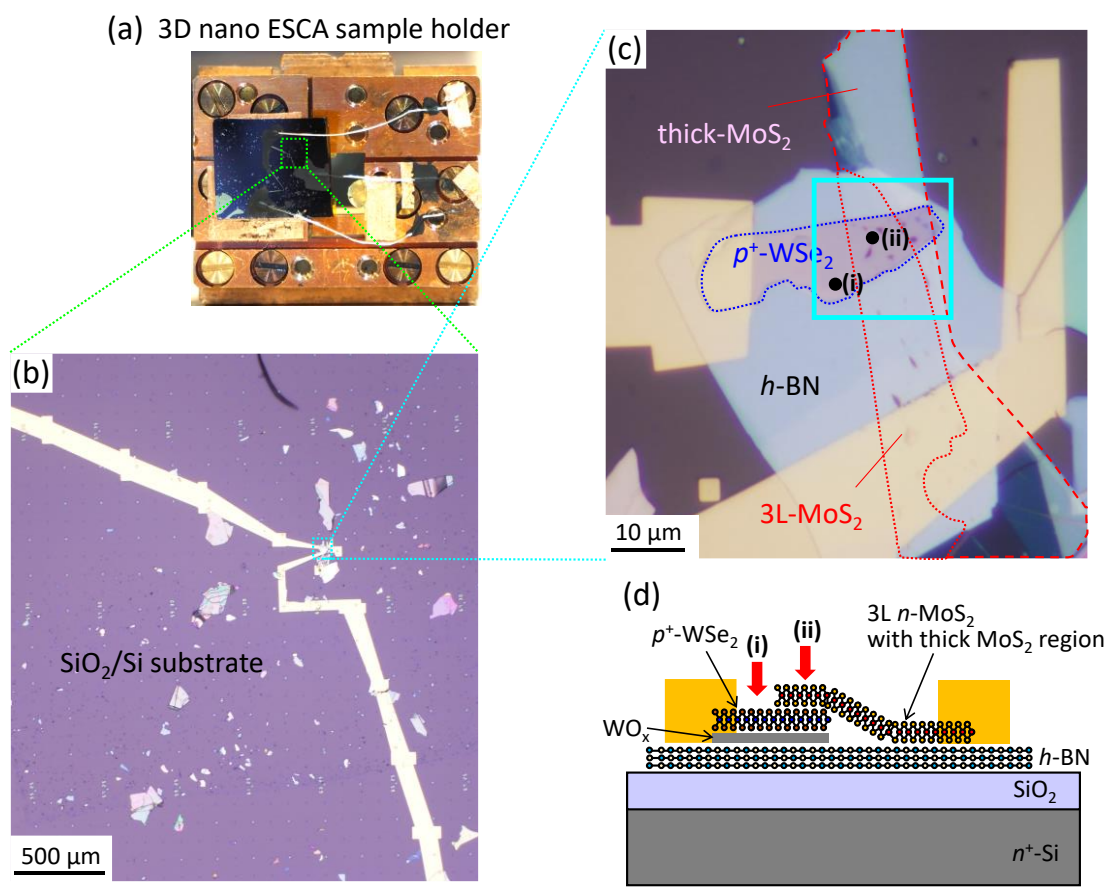

**Figure S4** 3D-nano ESCA installed at the University of Tokyo outstation beamline BL07LSU in SPring-8 used for chemical analysis. (a) Sample holder for the 3D-nano ESCA measurements, where the Ni/Au electrodes are grounded to the sample holder using Cu wire and carbon paste to prevent charge buildup on the $SiO_2$/Si substrate during the ESCA measurements. (b) Low magnification optical image of the device. (c) Magnified optical image of the 3L-$n$-$MoS_2$/$p^+$-$WSe_2$ heterostructure on $h$-BN with Ni/Au electrodes. (d) Schematic illustration of the heterostructure, showing the measurement points (i) and (ii).

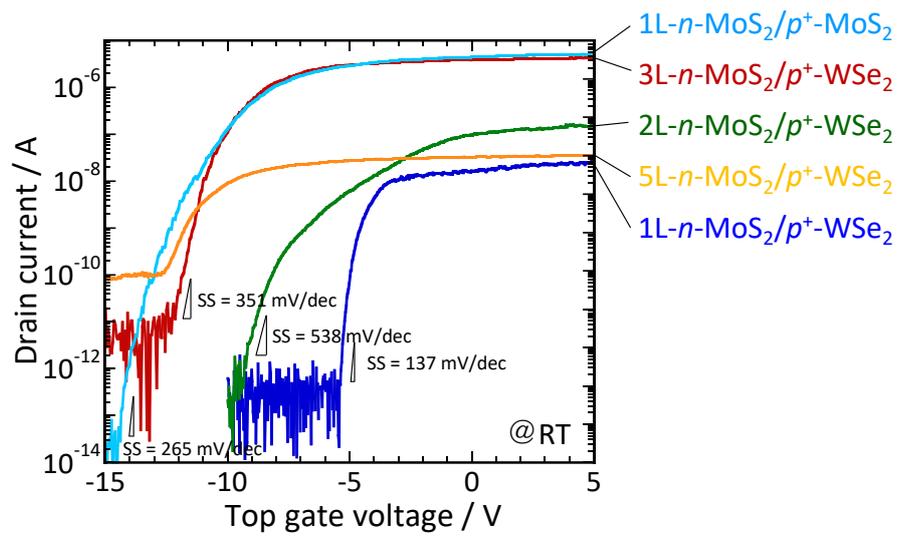

**Figure S5** Transfer characteristics of $n$-MoS$_2$/$p^+$-WSe$_2$ and $n$-MoS$_2$/$p^+$-MoS$_2$ TFETs with an ALD-Al$_2$O$_3$ top gate insulator. The minimum SS values are shown.

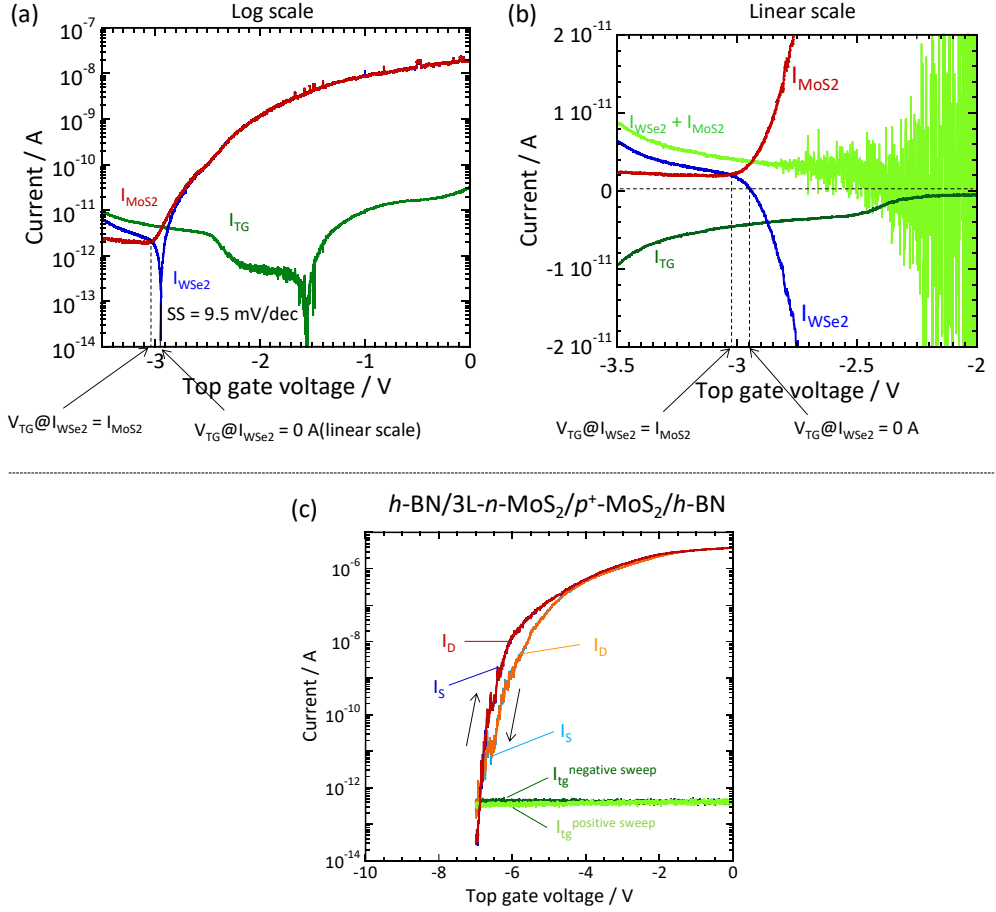

**Figure S6** (a) Transfer characteristics (log scale) of 1L-$n$-MoS$_2$/$p^+$-WSe$_2$ on the $h$-BN substrate with the $h$-BN top gate. The thicknesses of the top $h$-BN, $p^+$-WSe$_2$, and bottom $h$-BN are 7.5 nm, 43 nm, and 30 nm, respectively. (b) Linear scale of (a). (c) (c-1) Comparison of $I_D$ and $I_S$ for $h$-BN/3L-$n$-MoS$_2$/$p^+$-MoS$_2$/$h$-BN heterostructure TFET. (c-2) SS values for positive and negative sweeps.

**Note:** When $I_{MoS2}$ and $I_{WSe2}$ are compared in (a), SS = "9.5 mV/dec" in $I_{WSe2}$ is much smaller than that in $I_{MoS2}$. The origin for this artificially small SS value is discussed here. The linear scale is shown in (b), where $I_{TG}$ starts to flow at $V_{TG}$ = -2 V = $V_{WSe2}$ because the h-BN layer is thin. Because of this leakage, $I_{WSe2}$+$I_{MoS2}$ does not become zero, instead $I_{WSe2}$ + $I_{MoS2}$ = -$I_{TG}$, suggesting the contribution of $I_{TG}$ in $I_{WSe2}$. Here, $I_{WSe2}$ changes from the negative value to the positive value at $V_{TG}$@$I_{WSe2}$ = 0 A, which corresponds with the convex downward peak seen in (a). That is, the artificially small SS value for $I_{WSe2}$ results from the contribution of the leakage current. Unfortunately, this convex downward peak is quite often reported with the artificially small SS values in the previous literatures. To avoid this, for the $h$-BN/3L-$n$-MoS$_2$/$p^+$-MoS$_2$/$h$-BN heterostructure TFET, $I_D$ must be shown to be consistent with $I_S$, as shown in (c). The consistent data between $I_D$ and $I_S$ supports low SS value achieved here. In terms of the gate leakage $I_{TG}$ in (c), if there is no detectable gate current, $I_{TG}$ generally shows ~$10^{-13}$ A in our normal measurement setup (medium power SMU in Keysight B1500). Although this current level is higher than $I_D$ and $I_S$, this is typical case for no detectable gate leakage current. If there is small but detectable current, MPSMU can detect the range of $10^{-14}$ A or less, like $I_{TG}$ in (a). Therefore, the current level of ~$10^{-13}$ A indicates no gate leakage.

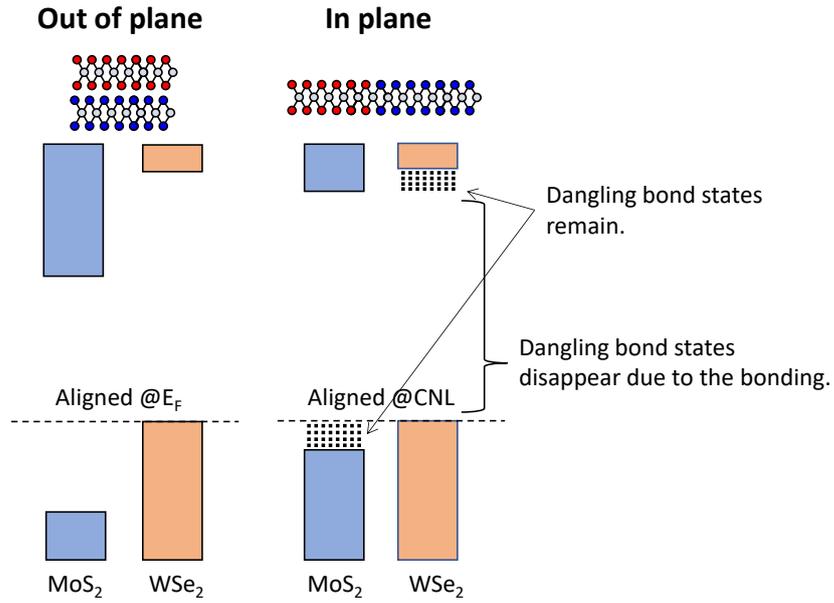

**Figure S7** Schematic comparison of band alignments for out-of-plane and in-plane $MoS_2/WSe_2$ junctions.[1] In the case of the out-of-plane structure, there is no dangling bond on the basal plane. Therefore, there is no dangling bond states in the interface. On the other hand, in case of in-plane structure, there is dangling bonds at the edge. When two 2D layers are ideally connected to each other, that is, there is no atomic defects at the interface, the dangling bond state remains at the position shown by the arrows. Therefore, out-of-plane heterostructures are preferable in terms of gate controllability.

Ref. [1] Y. Guo, J. Robertson, *Appl. Phys. Lett.* **2016**, *108*, 233104.